\documentclass[a4paper,11pt]{article}
\pdfoutput=1 

\usepackage{jheppub} 

\usepackage[T1]{fontenc} 
\usepackage{hyperref}

\title{Pentagon integrals to arbitrary order in the dimensional regulator}

\author[a,b]{Nikolaos Syrrakos}

\affiliation[a]{Institute of Nuclear and Particle Physics, NCSR "Demokritos",\\ Agia Paraskevi 15310, Greece}
\affiliation[b]{Physics Division, National Technical University of Athens, \\Athens 15780, Greece}

\emailAdd{syrrakos@inp.demokritos.gr}

\abstract{ We analytically calculate one-loop five-point Master Integrals, \textit{pentagon integrals}, with up to one off-shell leg to arbitrary order in the dimensional regulator in $d=4-2\epsilon$ space-time dimensions. A pure basis of Master Integrals is constructed for the pentagon family with one off-shell leg, satisfying a single-variable canonical differential equation in the Simplified Differential Equations approach. The relevant boundary terms are given in closed form, including a hypergeometric function which can be expanded to arbitrary order in the dimensional regulator using the \texttt{Mathematica} package \texttt{HypExp}. Thus one can obtain solutions of the canonical differential equation in terms of Goncharov Polylogartihms of arbitrary transcendental weight. As a special limit of the one-mass pentagon family, we obtain a fully analytic result for the massless pentagon family in terms of pure and universally transcendental functions. For both families we provide explicit solutions in terms of Goncharov Polylogartihms up to weight four.}


\begin{document} 
\maketitle
\flushbottom

\section{Introduction}
\label{sec:intro}
The study of one-loop five-point Feynman diagrams, known as pentagon integrals, has a long history. Their calculation has been treated using various methods and techniques, with results presented in several different forms. In \cite{Bern:1993kr}, analytic results for pentagon integrals with up to one massive leg were given in $d=4-2 \epsilon$ space-time dimensions in terms of transcendental functions of up to weight two. In \cite{DelDuca:2009ac}, the pentagon integral with fully massless legs was studied in $d=6-2 \epsilon$ space-time dimensions in the multi-Regge limit and analytic expressions up to $\mathcal{O}(\epsilon^2)$ were obtained. Analytic results in terms of Appell hypergeometric functions and Gauss hypergeometric functions were obtained in \cite{Kniehl:2010aj} for the pentagon with massless internal and external lines. First results in $d=4-2 \epsilon$ space-time dimensions in terms of Goncharov Polylogarithms \cite{Goncharov:1998kja} of up to weight three (or $\mathcal{O}(\epsilon)$) were obtained in \cite{Papadopoulos:2014lla} for the pentagon with up to one off-shell leg. One-fold integral representations were obtained in \cite{Kozlov:2015kol} in arbitrary space-time dimensions for the massless pentagon and in \cite{Kozlov:2016vqy} for the pentagon with one off-shell leg in $d=6-2 \epsilon$ space-time dimensions. More recently the massless pentagon has been implemented as part of the so-called \textit{pentagon functions} up to weight four in \cite{Gehrmann:2018yef, Chicherin:2020oor} and numerical results up to weight four in terms of generalised power-series expansion for the pentagon with one off-shell leg were presented in \cite{Abreu:2020jxa} for various phase-space points. 

In this paper we analytically calculate pentagon integrals with up to one off-shell leg in $d=4-2 \epsilon$ space-time dimensions and express them in terms of Goncharov Polylogarithms of higher weights. We start with the pentagon with one massive leg and employ the method of differential equations \cite{de1, de2, de3, de4} in its modern, canonical incarnation \cite{Henn:2013pwa}, in conjunction with the Simplified Differential Equations approach (SDE) \cite{Papadopoulos:2014lla}. This is achieved by constructing a \textit{pure basis} of master integrals \cite{Henn:2014qga} for the pentagon with one massive leg family. We find that all relevant boundary terms required for the solution of the canonical differential equation are given in closed form. This allows us to straightforwardly express our results in terms of Goncharov Polylogarithms of arbitrary transcendental weight. Having the solution of the pentagon with one off-shell leg at hand, we can obtain in an algorithmic way a fully analytical solution for the massless pentagon by taking the $x\to1$ limit of the former result, with $x$ being a dimensionless parameter that is introduced in the SDE approach. This is a special feature of the SDE approach first described in \cite{Papadopoulos:2014lla, Papadopoulos:2015jft} and in more details in \cite{Canko:2020gqp}.

With a mind to phenomenological applications of our results, we present explicit solutions up to transcendental weight four. Our results are complementary to \cite{Canko:2020ylt}, where analytic expressions for all two-loop planar pentabox families with one massive leg were recently presented. The combination of these results provides a key ingredient for the calculation of two-loop planar amplitudes for W+2 jets production at the LHC \cite{Hartanto:2019uvl, Badger:2021nhg}.

Our paper is structured as follows: in section  \ref{sec:basis} we introduce basic notation and construct a pure basis of Master Integrals for the pentagon family with one off-shell leg. In section  \ref{sec:Result} we compute all relevant boundary terms and provide solutions in terms of Goncharov Polylogarithms up to weight four. In section  \ref{sec:Massless} we obtain a solution for the massless pentagon family through the $x\to1$ limit of the pentagon family with one off-shell leg solution. In section \ref{sec:Checks} we report on the numerical checks of our solutions for both Euclidean and physical points. Finally, section \ref{sec:conclusions} provides a brief summary of our findings. For convenience we provide appendices with the alphabet, the pure basis for the one-mass pentagon family, as well as the boundary terms in closed form.

\section{Construction of a pure basis}
\label{sec:basis}

We define the pentagon family with one off-shell leg as 

\begin{equation}\label{eq:pentaf}
    G_{a_1a_2a_3a_4a_5} = \int~ \frac{\mathrm{d}^d k_1}{i\pi^{(d/2)}}~ \frac{\mathrm{e}^{\epsilon \gamma_{E}}}{\mathcal{D}_1^{a_1}\mathcal{D}_2^{a_2}\mathcal{D}_3^{a_3}\mathcal{D}_4^{a_4}\mathcal{D}_5^{a_5}}
\end{equation}
with
\begin{align}
    &\mathcal{D}_1=-(k_1)^2,~ \mathcal{D}_2=-(k_1+q_1)^2,~ \mathcal{D}_3=-(k_1+q_1+q_2)^2 \nonumber \\
    &\mathcal{D}_4=-(k_1+q_1+q_2+q_3)^2,~ \mathcal{D}_5=-(k_1+q_1+q_2+q_3+q_4)^2
\end{align}
The kinematics consist of the massive momentum $\{q_3,~ q_3^2=m^2\}$ and four massless momenta $\{q_i,~ q_i^2=0$, for $i=1,2,4,5\}$. In the SDE notation, we parametrise the external momenta by inserting a dimensionless variable $x$ in the following way\footnote{We use the abbreviations $p_{ij}=p_i+p_j$ and $p_{ijk}=p_i+p_j+p_k$ and similarly for $q$ later.}
\begin{equation}
    q_1=xp_1,~q_2=xp_2,~q_3=p_{123}-xp_{12},~q_{4}=-p_{123},~q_5=-p_{1234}
\end{equation}
where all $p_i$ momenta are now massless. The kinematics of the two momentum configurations are connected as follows\footnote{We use the abbreviations $s_{ij}=q_{ij}^2,~ S_{ij}=p_{ij}^2$.}
\begin{align}
    &m^2 = (x-1) \left(S_{12} x-S_{45}\right),~s_{12} = S_{12} x^2,~s_{23} = S_{23} x-S_{45} x+S_{45},\nonumber \\
    &s_{34} = x \left(S_{12} (x-1)+S_{34}\right),~s_{45} = S_{45},~s_{51} = S_{51} x
\end{align}
For reasons that will become clear shortly, we also define the square root of the Gram determinant of the external momenta as
\begin{equation}
    \Delta_5 = \sqrt{\det[q_i\cdot q_j]}
\end{equation}

Using \texttt{FIRE6} \cite{Smirnov:2019qkx} for the IBP reduction \cite{IBP1, IBP2} we find that we need 13 Master Integrals to completely describe the vector space which consists of all Feynman Integrals that are included in \eqref{eq:pentaf}. For the construction of the pure basis, the only non-trivial basis element is that of the top sector. The remaining basis elements can be constructed through the knowledge of the leading singularities of the relevant one-loop integrals. For the construction of the top sector basis element we follow the consensus developed in \cite{Abreu:2018rcw, Abreu:2020jxa}, expressed here in the language of the Baikov representation of Feynman Integrals \cite{Baikov:1996iu, Baikov:1996rk}. More specifically, we choose as a candidate top sector basis element the following
\begin{equation}\label{eq:utpenta}
    \epsilon^2 \frac{\mathcal{P}_{11111}}{\Delta_5} G_{11111}
\end{equation}
where $\mathcal{P}_{11111}$ is the Baikov polynomial of $G_{11111}$. By studying the maximal cut of \eqref{eq:utpenta} in the Baikov represenation \cite{Frellesvig:2017aai}, we can see that it has the desired properties of being pure and universally transcendental. This result is a strong indicator that the uncut \eqref{eq:utpenta} will satisfy a canonical differential equation. This is due to the fact that as shown in \cite{Henn:2014qga}, since cut and uncut integrals satisfy the same system of IBP identities and hence the same system of differential equations, if one is looking for a system of canonical differential equations, then the same canonical differential equations have to be satisfied both by the cut and the uncut integrals. Therefore, a pure master integral of uniform weight that satisfies a canonical differential equation, has to preserve these properties on its maximal cut. Thus, when one has identified an integral, or a combination of integrals, which in its maximal cut is expressed in terms of pure functions of uniform weight, then this integral, or combination of integrals, is a strong candidate for an element of the desired pure basis. 

In \cite{Abreu:2020jxa}, a pure basis for the pentagon family with one off-shell leg is given in the ancillary files. The top sector basis element \eqref{eq:utpenta} is the same as the one given in \cite{Abreu:2020jxa}, albeit in a different notation. 

For convenience we choose to express our pure basis in terms of specific Feynman Integrals in the form
\begin{equation}\label{eq:gtg}
    \textbf{g}=\textbf{T}\textbf{G}
\end{equation}
where $\textbf{g}$ is the pure basis, \textbf{G} is our specific choice of Feynman Integrals and $\textbf{T}$ is the matrix that connects the two bases. Both the pure basis $\textbf{g}$  as well as our choice of Feynman Integrals \textbf{G} are given in appendix \ref{sec:PureB}. 

\section{Results}
\label{sec:Result}

The ultimate check that a candidate pure basis of Master Integrals has the desired properties is the derivation of its differential equation \cite{Henn:2013pwa}. In the Simplified Differential Equations approach \cite{Papadopoulos:2014lla}, we derive differential equations with respect to the dimensionless parameter $x$, regardless of the number of scales of the problem. In this particular case, the candidate pure basis which was constructed in the previous section leads to the following canonical differential equation
\begin{equation}\label{eq:cande}
\partial_{x} \textbf{g}=\epsilon \left( \sum_{i=1}^{11} \frac{\textbf{M}_i}{x-l_i} \right) \textbf{g}
\end{equation}
where $\textbf{g}$ is the array of 13 pure Master Integrals, $\textbf{M}_i$ are the residue matrices which consist solely of rational numbers and $l_i$ are the letters of the alphabet. In appendix \ref{sec:Letters} we present the letters of the alphabet for this particular family. Notice that here we follow \cite{Papadopoulos:2015jft, Canko:2020gqp, Canko:2020ylt} for the definition of the letters of the alphabet, which is different from the standard notation \cite{Duhr:2011zq, Duhr:2012fh, Duhr:2014woa}. Usually the so-called $\mathrm{d}\log$ form of a system of canonical differential equations is given as $\mathrm{d}\textbf{g}(\vec{x}, \epsilon) = \epsilon \bigg(\sum_{i} \textbf{M}_i \mathrm{d}\log W_i (\vec{x}) \bigg) \textbf{g}(\vec{x}, \epsilon)$, where the alphabet $W_i (\vec{x})$ is in terms of rational or algebraic functions of the independent variables. The standard $\mathrm{d}\log$ form is equivalent to \eqref{eq:cande} for $W_i (\vec{x}) = x-l_i$.

This differential equation can be solved through recursive iterations up to the desired order in the dimensional regulator $\epsilon$. The only further input that is required are the boundary terms. In the Simplified Differential Equations approach, we choose as a lower integration boundary $x=0$, so that the result can be expressed directly in terms of Goncharov Polylogarithms. Thus the necessary boundary terms are given by the limit $x\to0$ of our chosen basis of Master Integrals.

Using the methods described in \cite{Canko:2020ylt, Canko:2020gqp}, we define the \textit{resummation matrix} $\textbf{R}$ as follows
\begin{equation}\label{eq:resm0}
    \textbf{R} =\textbf{S} e^{\epsilon \textbf{D} \log(x)} \textbf{S}^{-1}
\end{equation}
where $\textbf{D},~\textbf{S}$ are defined through the Jordan decomposition\footnote{For an earlier use of the Jordan decomposition method see also \cite{Dulat:2014mda, Mastrolia:2017pfy}. We thank Prof. P. Mastrolia for bringing these references to our attention.} of the residue matrix corresponding to the letter $\{0\}$, $\textbf{M}_1$,
\begin{equation}
    \textbf{M}_1 =\textbf{S} \textbf{D} \textbf{S}^{-1}
\end{equation}
Furthermore, using the expansion-by-regions method \cite{Jantzen:2012mw} we obtain information for the $x\to0$ limit of the Feynman Integrals in terms of which we express the pure basis of Master Integrals \eqref{eq:gtg},
\begin{equation}\label{eq:regions}
    {G_i}\mathop  = \limits_{x \to 0} \sum\limits_j x^{b_j + a_j \epsilon }G^{(b_j + a_j \epsilon)}_{i} 
\end{equation}
where $a_j$ and $b_j$ are integers and $G_i$ are the individual members of the basis $\textbf{G}$ of Feynman Integrals in \eqref{eq:gtg}, by making use of the publicly available \texttt{FIESTA4}~\cite{Smirnov:2015mct} code. As explained in \cite{Canko:2020ylt}, we can construct the relation
\begin{equation}\label{eq:bounds}
    \mathbf{R} \mathbf{b}=\left.\lim _{x \rightarrow 0} \mathbf{T} \mathbf{G}\right|_{\mathcal{O}\left(x^{0+a_{j} \epsilon}\right)}
\end{equation}
where $\textbf{b}=\sum_{i=0}^n\epsilon^i\textbf{b}_0^{(i)}$ are the boundary terms that we need to compute. The right-hand-side of  \eqref{eq:bounds} implies that, apart from the terms $x^{a_i  \epsilon}$ coming from \eqref{eq:regions}, we expand around $x=0$, keeping only terms of order $x^0$. In contrast with \cite{Canko:2020ylt}, in this case no logarithmic terms arise in the left-hand-side of  \eqref{eq:bounds}, i.e. the matrix $\textbf{D}$ in \eqref{eq:resm0} is a diagonal matrix. From \eqref{eq:bounds} we can straightforwardly obtain all but one boundary terms without any further computation. This is due to the fact that all but one boundary terms are either given in terms of two-point functions or zero. The seventh basis element requires the computation of the leading region for $G_{11101}$ which corresponds to $G_{7}^{-3-2\epsilon}$ in the notation of \eqref{eq:regions}. Through its Feynman parametrization this region can be computed in terms of a hypergeometric function and the final result for the boundary term of $g_7$ is 
\begin{equation}
   \mathrm{b}_7 = -\frac{e^{\gamma_{E}  \epsilon } \epsilon  \left(-S_{12}\right){}^{-\epsilon } \Gamma (-\epsilon )^2 \Gamma (\epsilon +1) \, _2F_1\left(1,-\epsilon ;1-\epsilon ;\frac{S_{12}-S_{34}+S_{51}}{S_{51}}\right)}{\Gamma (-2 \epsilon )}
\end{equation}
In appendix \ref{sec:Bounds} we present all the necessary boundary terms in closed form. Using standard algorithms as implemented in modern computer algebra systems, e.g. \texttt{Mathematica}, and with the additional help of the public package \texttt{HypExp}~\cite{Huber:2007dx} for the series expansion of the hypergeometric function, we can obtain exact expressions for all boundary terms up to arbitrary order in $\epsilon$. Therefore we can trivially obtain solutions of \eqref{eq:cande} in terms of Goncharov Polylogarithms up to arbitrary weight.

In this particular work, we present explicit results up to weight four, which can be written in compact form as
\begin{align}
   \label{eq:solution}
   \textbf{g}&= \epsilon^0 \textbf{b}^{(0)}_{0} + \epsilon \bigg(\sum\mathcal{G}_{a}\textbf{M}_{a}\textbf{b}^{(0)}_{0}+\textbf{b}^{(1)}_{0}\bigg) \nonumber \\
   &+ \epsilon^2 \bigg(\sum\mathcal{G}_{ab}\textbf{M}_{a}\textbf{M}_{b}\textbf{b}^{(0)}_{0}+\sum\mathcal{G}_{a}\textbf{M}_{a}\textbf{b}^{(1)}_{0}+\textbf{b}^{(2)}_{0}\bigg) \nonumber \\
   &+ \epsilon^3 \bigg(\sum\mathcal{G}_{abc}\textbf{M}_{a}\textbf{M}_{b}\textbf{M}_{c}\textbf{b}^{(0)}_{0}+\sum\mathcal{G}_{ab}\textbf{M}_{a}\textbf{M}_{b}\textbf{b}^{(1)}_{0}+\sum\mathcal{G}_{a}\textbf{M}_{a}\textbf{b}^{(2)}_{0}+\textbf{b}^{(3)}_{0}\bigg) \nonumber \\
   &+ \epsilon^4 \bigg(\sum\mathcal{G}_{abcd}\textbf{M}_{a}\textbf{M}_{b}\textbf{M}_{c}\textbf{M}_{d}\textbf{b}^{(0)}_{0}+\sum\mathcal{G}_{abc}\textbf{M}_{a}\textbf{M}_{b}\textbf{M}_{c}\textbf{b}^{(1)}_{0}\nonumber \\
   &+ \sum\mathcal{G}_{ab}\textbf{M}_{a}\textbf{M}_{b}\textbf{b}^{(2)}_{0}+\sum\mathcal{G}_{a}\textbf{M}_{a}\textbf{b}^{(3)}_{0}+\textbf{b}^{(4)}_{0}\bigg)
   \\
   \mathcal{G}_{ab\ldots}&:= \mathcal{G}(l_a,l_b,\ldots;x) \nonumber
\end{align}
with the generalisation to higher orders in $\epsilon$ being straightforward.

\section{Massless pentagon family}
\label{sec:Massless}

Using the methods described in \cite{Canko:2020gqp}, we can readily obtain a pure basis of 11 Master Integrals for the massless pentagon family from the $x\to1$ limit of \eqref{eq:solution}. The results are by construction in terms of Goncharov Polylogarithms up to weight four, however following the arguments of the last section, we can obtain results in terms of Goncharov Polylogarithms of arbitrary weight.

To be more specific, as described in \cite{Canko:2020gqp} the $x\to1$ limit of the solution for the pentagon with one off-shell leg can be given by the following formula
\begin{equation}\label{eq:lim1}
    \textbf{g}_{x\to1}=\Tilde{\textbf{R}}_0\textbf{g}_{trunc}
\end{equation}
with $\Tilde{\textbf{R}}_0$ being a purely numerical matrix which is constructed from the resummation matrix 
\begin{equation}\label{eq:resum1}
    \Tilde{\textbf{R}}  = \Tilde{\textbf{S}} e^{\epsilon \Tilde{\textbf{D}} \log(1-x)} \Tilde{\textbf{S}}^{-1}
\end{equation}
after setting all terms of $(1-x)^{a_i\epsilon}$ in $\Tilde{\textbf{R}}$ equal to zero\footnote{Here $a_i=\{-1,0\}$ are the eigenvalues of $\textbf{M}_4$.}. The matrices $\Tilde{\textbf{S}}, \Tilde{\textbf{D}}$ are constructed through the Jordan decomposition of the residue matrix corresponding to the letter $\{1\}$, which in this case is $\textbf{M}_4$, i.e. $\textbf{M}_4 = \Tilde{\textbf{S}} \Tilde{\textbf{D}} \Tilde{\textbf{S}}^{-1}$.

The second input in \eqref{eq:lim1}, $\textbf{g}_{trunc}$, is just the regular part of the solution \eqref{eq:solution} at $x\to1$ where we have set $x=1$ explicitly \cite{Canko:2020gqp}.

\section{Numerical checks}
\label{sec:Checks}
We have performed various numerical checks of our results. Firstly, for the pentagon family with one leg off-shell we compared our results with \texttt{pySecDec}~\cite{Borowka:2017idc} for the Euclidean point
\begin{equation}
    S_{12}\to -2,~S_{23}\to -3,~S_{34}\to -5,~S_{45}\to -7,~S_{51}\to -11,~x\to \frac{1}{4}
\end{equation}
For numerical results in physical regions, we refer to the detailed discussion in section 4 of \cite{Canko:2020ylt}. Using the methods described therein, we managed to analytically continue our results for the pentagon family with one-off shell leg and compare with the numerical results of \cite{Abreu:2020jxa} for all the physical points that are provided there.

For the massless pentagon family, we have performed numerical checks against \texttt{pySecDec}~\cite{Borowka:2017idc} for the Euclidean point
\begin{equation}
    S_{12}\to -2,~S_{23}\to -3,~S_{34}\to -5,~S_{45}\to -7,~S_{51}\to -11
\end{equation}

For the numerical evaluation of the relevant Goncharov Polylogarithms, we utilised \texttt{Ginac}~\cite{Vollinga:2005pk, Vollinga:2004sn} through the \texttt{Mathematica} interface provided by \texttt{PolyLogTools}~\cite{Duhr:2019tlz}. The latter package was also used extensively for the manipulation of the resulting Goncharov Polylogarithms. For all checks that were carried out we report excellent agreement.

\section{Conclusions}
\label{sec:conclusions}

In this paper we have presented the analytic calculation of pentagon integrals with up to one off-shell leg in $d=4-2 \epsilon$ space-time dimensions in terms of Goncharov Polylogarithms of higher weights. We constructed a candidate pure basis of master integrals for the pentagon family with one off-shell leg following the ideas of \cite{Abreu:2018rcw, Abreu:2020jxa} and we verified that it satisfies a canonical differential equation \cite{Henn:2013pwa} in the Simplified Differential Equations approach \cite{Papadopoulos:2014lla}.

We have demonstrated that all relevant boundary terms are either zero or given in closed form. This allows one to straightforwardly obtain analytic solutions of the canonical differential equation for the pentagon family with one off-shell leg in terms of Goncharov Polylogarithms \cite{Goncharov:1998kja} of arbitrary transcendental weight. As a by-product of this result, we were able to obtain analytic expressions for the massless pentagon family in terms of Goncharov Polylogarithms, as well as a pure basis of master integrals, after taking the $x\to 1$ limit of the former result in an algorithmic way \cite{Canko:2020gqp}.

For both families explicit results in terms of Goncharov Polylogarithms of up to weight four are provided. Our results are complementary to the recently published analytic expressions for the two-loop pentaboxes with one off-shell leg \cite{Canko:2020ylt} and can be used for the calculation of two-loop planar amplitudes for W+2 jets production \cite{Hartanto:2019uvl, Badger:2021nhg}. We have also demonstrated the validity of our solutions for both Euclidean and physical points. 

Finally, all the relevant results of our analysis can be found in
\begin{center}
    \href{https://github.com/nsyrrakos/Pentagon.git}{https://github.com/nsyrrakos/Pentagon.git}.
\end{center}
We provide the pure bases for both pentagon families (\textbf{Basis} for the off-shell leg family and \textbf{Basis0} for the massless one), solutions for both families up to weight four (\textbf{Pentagonw4} for the off-shell leg family and \textbf{Pentagon0w4} for the massless one), the residue matrices as well as the letters for the canonical differential equation \eqref{eq:cande} (\textbf{DE\textunderscore Matrices\textunderscore letters}), as well as the necessary boundary terms in closed form (\textbf{Boundaries}). Finally, we provide the specific choice of Feynman Integrals $\textbf{G}$ in \eqref{eq:gtg} ($\textbf{Masters}$).

\appendix

\section{Alphabet}
\label{sec:Letters}
The alphabet of the pentagon family with one off-shell leg has the following 11 letters,
\begin{align}
    &l_1\to 0,l_2\to \frac{S_{12}-S_{34}}{S_{12}},l_3\to \frac{S_{45}}{S_{45}-S_{23}},l_4\to 1,l_5\to \frac{S_{45}}{S_{12}},l_6\to \frac{S_{12}-S_{34}+S_{51}}{S_{12}}, \nonumber \\ 
    &l_7\to \frac{S_{45}}{-S_{23}+S_{45}+S_{51}},l_8\to \frac{S_{45}}{S_{34}+S_{45}},l_9\to \frac{S_{12}+S_{23}}{S_{12}}, \nonumber \\ 
    &l_{10}\to \frac{\sqrt{\Delta }+S_{12} S_{23}-S_{23} S_{34}-2 S_{12} S_{45}+S_{34} S_{45}-S_{12} S_{51}-S_{45} S_{51}}{2 S_{12} \left(S_{23}-S_{45}-S_{51}\right)}, \nonumber \\ 
    &l_{11}\to \frac{-\sqrt{\Delta }+S_{12} S_{23}-S_{23} S_{34}-2 S_{12} S_{45}+S_{34} S_{45}-S_{12} S_{51}-S_{45} S_{51}}{2 S_{12} \left(S_{23}-S_{45}-S_{51}\right)}
\end{align}
with $\Delta$ being
\begin{align}
    \Delta &= \left(S_{23} S_{34}+S_{45} \left(S_{51}-S_{34}\right)+S_{12} \left(-S_{23}+2 S_{45}+S_{51}\right)\right){}^2 \nonumber \\
    &-4 S_{12} S_{45} \left(S_{12}-S_{34}+S_{51}\right) \left(-S_{23}+S_{45}+S_{51}\right)
\end{align}

\section{Pure basis}
\label{sec:PureB}

\begin{align}
g_1 &= x \epsilon  S_{51} G_{0,2,0,0,1}\\
g_2 &= x \epsilon  \left((x-1) S_{12}+S_{34}\right) G_{0,0,2,0,1}\\
g_3 &= x^2 \epsilon  S_{12} G_{2,0,1,0,0}\\
g_4 &= \epsilon  S_{45} G_{2,0,0,1,0}\\
g_5 &= \epsilon  \left(x S_{23}-x S_{45}+S_{45}\right) G_{0,2,0,1,0}\\
g_6 &= (x-1) \epsilon  \left(x S_{12}-S_{45}\right) G_{0,0,2,1,0}\\
g_7 &= x^3 \epsilon ^2 S_{12} S_{51} G_{1,1,1,0,1}\\
g_8 &= x \epsilon ^2 S_{45} S_{51} G_{1,1,0,1,1}\\
g_9 &= x \epsilon ^2 \left((x-1) S_{12}+S_{34}\right) S_{45} G_{1,0,1,1,1}\\
g_{10} &= x \epsilon ^2 \big(-x S_{23} S_{34}+(x-1) S_{45} S_{34}-(x-1) S_{45} S_{51} \nonumber \\
&+(x-1) S_{12} \left(x \left(-S_{23}+S_{45}+S_{51}\right)-S_{45}\right)\big) G_{0,1,1,1,1}\\
g_{11} &= x \epsilon ^2 \left(S_{12}-S_{45}\right) G_{1,0,1,1,0}\\
g_{12} &= x^2 \epsilon ^2 S_{12} \left(x S_{23}-x S_{45}+S_{45}\right) G_{1,1,1,1,0}\\
g_{13} &= \frac{x \epsilon^2}{32 \sqrt{\hat{\Delta}}} \bigg(A_1 G_{0,1,1,1,1} + A_2 G_{1,0,1,1,1} + A_3 G_{1,1,0,1,1} + A_4 G_{1,1,1,0,1} + A_5 G_{1,1,1,1,0} + A_6 G_{1,1,1,1,1}\bigg)
\end{align}
with 
\begin{align}
\hat{\Delta} &= S_{12}^2 \left(S_{23}-S_{51}\right){}^2 + \left(S_{23} S_{34}+S_{45} \left(S_{51}-S_{34}\right)\right){}^2 \nonumber \\
&+ 2 S_{12} \left(S_{34} S_{45} S_{23}+\left(S_{34}+S_{45}\right) S_{51} S_{23}-S_{23}^2 S_{34}+S_{45} \left(S_{34}-S_{51}\right) S_{51}\right)
\end{align}
and
\begin{align}
A_1 &= \left(S_{12} S_{23}-S_{34} S_{23}+S_{34} S_{45}-\left(S_{12}+S_{45}\right) S_{51}\right) \nonumber \\
& \times \left(S_{34} \left(S_{23} x-S_{45} x+S_{45}\right)+S_{45} S_{51} (x-1)-S_{12} (x-1) \left(\left(-S_{23}+S_{45}+S_{51}\right) x-S_{45}\right)\right)\\
A_2 &= S_{45} \big(S_{34} \left(S_{23} S_{34}+S_{45} \left(S_{51}-S_{34}\right)\right)-S_{12}^2 \left(S_{23}-S_{51}\right) (x-1)\nonumber \\
& + S_{12} \left(S_{23} S_{34} (x-2)+S_{34} \left(-\left(S_{45}+2 S_{51}\right) x+S_{45}+S_{51}\right)-S_{45} S_{51} (x-1)\right) \big)\\
A_3 &= -S_{45} S_{51} \left(S_{12} \left(S_{23} (2 x-1)-2 \left(S_{45}+S_{51}\right) x+2 S_{45}+S_{51}\right)+S_{23} S_{34}+S_{45} \left(S_{51}-S_{34}\right)\right)\\
A_4 &= S_{12} S_{51} (-x) \left(S_{23} S_{34} x-S_{45} S_{34} (x-2)+2 S_{12} S_{45} (x-1)+S_{45} S_{51} (x-2)+S_{12} \left(S_{51}-S_{23}\right) x\right)\\
A_5 &= S_{12} x \big( S_{34} S_{23}^2 x+S_{45}^2 \left(S_{34}-S_{51}\right) (x-1) + S_{23} S_{45} \left(-2 S_{34} x+S_{51} (x-2)+S_{34}\right)\nonumber \\
& +S_{12} \left(S_{23}^2 (-x)+\left(S_{45}+S_{51}\right) S_{23} x+S_{45} S_{51} (x-1)-S_{45} S_{23}\right)  \big)\\
A_6 &= -2 S_{12} S_{45} S_{51} x \left(S_{34} \left(S_{23} x-S_{45} x+S_{45}\right)+S_{45} S_{51} (x-1)-S_{12} (x-1) \left(\left(-S_{23}+S_{45}+S_{51}\right) x-S_{45}\right)\right)
\end{align}

The particular choice of Feynman Integrals in terms of which the above pure basis is written is the following
\begin{align}
\textbf{G}=\big\{&G_{0,2,0,0,1},G_{0,0,2,0,1},G_{2,0,1,0,0},G_{2,0,0,1,0},G_{0,2,0,1,0},G_{0,0,2,1,0}, \nonumber \\
&G_{1,1,1,0,1},G_{1,1,0,1,1},G_{1,0,1,1,1},G_{0,1,1,1,1},G_{1,0,1,1,0},G_{1,1,1,1,0},G_{1,1,1,1,1}\big\}
\end{align}

\section{Boundary terms in closed form}
\label{sec:Bounds}

\begin{align}
   &\text{b}_1=\frac{e^{\gamma_{E}  \epsilon } \left(-S_{51}\right){}^{-\epsilon } \Gamma (1-\epsilon )^2 \Gamma (\epsilon +1)}{\Gamma (1-2 \epsilon )} \\
   &\text{b}_2=\frac{e^{\gamma_{E}  \epsilon } \left(S_{12}-S_{34}\right){}^{-\epsilon } \Gamma (1-\epsilon )^2 \Gamma (\epsilon +1)}{\Gamma (1-2 \epsilon )} \\
   &\text{b}_3=\frac{e^{\gamma_{E}  \epsilon } \left(-S_{12}\right){}^{-\epsilon } \Gamma (1-\epsilon )^2 \Gamma (\epsilon +1)}{\Gamma (1-2 \epsilon )} \\
   &\text{b}_4=\frac{e^{\gamma_{E}  \epsilon } \left(-S_{45}\right){}^{-\epsilon } \Gamma (1-\epsilon )^2 \Gamma (\epsilon +1)}{\Gamma (1-2 \epsilon )} \\
   &\text{b}_5=\frac{e^{\gamma_{E}  \epsilon } \left(-S_{45}\right){}^{-\epsilon } \Gamma (1-\epsilon )^2 \Gamma (\epsilon +1)}{\Gamma (1-2 \epsilon )} \\
   &\text{b}_6=\frac{e^{\gamma_{E}  \epsilon } \left(-S_{45}\right){}^{-\epsilon } \Gamma (1-\epsilon )^2 \Gamma (\epsilon +1)}{\Gamma (1-2 \epsilon )} \\
   &\text{b}_7=-\frac{e^{\gamma_{E}  \epsilon } \epsilon  \left(-S_{12}\right){}^{-\epsilon } \Gamma (-\epsilon )^2 \Gamma (\epsilon +1) \, _2F_1\left(1,-\epsilon ;1-\epsilon ;\frac{S_{12}-S_{34}+S_{51}}{S_{51}}\right)}{\Gamma (-2 \epsilon )} \\
   &\text{b}_8=\frac{2 e^{\gamma_{E}  \epsilon } \left(-S_{51}\right){}^{-\epsilon } \Gamma (1-\epsilon )^2 \Gamma (\epsilon +1)}{\Gamma (1-2 \epsilon )} \\
   &\text{b}_9=\frac{e^{\gamma_{E}  \epsilon } \left(-S_{12}\right){}^{-\epsilon } \left(2 \left(-S_{12}\right){}^{\epsilon }-\left(S_{12}-S_{34}\right){}^{\epsilon }\right) \left(S_{12}-S_{34}\right){}^{-\epsilon } \Gamma (1-\epsilon )^2 \Gamma (\epsilon +1)}{\Gamma (1-2 \epsilon )} \\
   &\text{b}_{10}=\frac{2 e^{\gamma_{E}  \epsilon } \left(S_{12}-S_{34}\right){}^{-\epsilon } \left(\left(S_{12}-S_{34}\right){}^{\epsilon }-\left(-S_{51}\right){}^{\epsilon }\right) \left(-S_{51}\right){}^{-\epsilon } \Gamma (1-\epsilon )^2 \Gamma (\epsilon +1)}{\Gamma (1-2 \epsilon )} \\
   &\text{b}_{11}=0 \\
   &\text{b}_{12}=\frac{e^{\gamma_{E}  \epsilon } \left(-S_{12}\right){}^{-\epsilon } \Gamma (1-\epsilon )^2 \Gamma (\epsilon +1)}{\Gamma (1-2 \epsilon )} \\ 
   &\text{b}_{13}=0
\end{align}

\acknowledgments
The author would like to thank Costas G. Papadopoulos and Dhimiter D. Canko for valuable discussions.


\end{document}